\documentclass[10pt]{elsart}

\usepackage{graphicx}

\usepackage{amssymb}

\def\dt{\Delta t}
\def\dx{\Delta x}
\def\p{\partial}
\def\grad{\nabla}
\def\vx{{\bf x}}
\def\vk{{\bf k}}

\begin{document}

\begin{frontmatter}



\title{An Efficient Algorithm for Solving the Phase Field Crystal Model}
\author{Mowei Cheng\corauthref{cor}} and
\corauth[cor]{Corresponding author. Tel: +1-301-975-5729; Fax: +1-301-975-5012.}
\ead{mowei.cheng@nist.gov}
\author{James A. Warren}
\address{Metallurgy Division and Center for Theoretical and Computational Materials
Science, National Institute of Standards and Technology, 100 Bureau Drive, Stop 8554,
Gaithersburg, Maryland 20899, USA}

\begin{abstract}
  
  We present and discuss the development of an unconditionally stable
  algorithm used to solve  the evolution equations of the Phase Field
  Crystal (PFC) model. This algorithm allows for an arbitrarily large
  algorithmic time step. As the basis for our analysis of the accuracy
  of this algorithm, we determine an effective time step in Fourier
  space. We then compare our calculations with a set of representative
  numerical results, and demonstrate that this algorithm is an
  effective approach for the study of the PFC models, yielding a time
  step effectively 180 times larger than the Euler algorithm for a representative
  set of material parameters.  As the PFC model is just a simple
  example of a wide class of density functional theories, we expect
  this method will have wide applicability to modeling systems of
  considerable interest to the materials modeling communities.

\end{abstract}



\end{frontmatter}


\section{Introduction}

The dynamics of a non-equilibrium system often results in highly
complicated domain structures (microstructures).  Typically, as time
proceeds, the average size of these structures grows as a direct
consequence of free-energy reduction: the interface is eliminated resulting
in an increase in the size of homogeneous regions.  Traditional
non-equilibrium dynamics usually deals with the equilibrium states
that are spatially uniform \cite{Langer1,Bray1,Gunton1,Furukawa1},
i.e., the stable phases are characterized by homogeneous values for the
appropriate intensive thermodynamic variables.  Classic, albeit quite
simple, examples of models governing the evolution of such systems are
the Cahn-Hilliard (CH) Equation for conserved systems \cite{Cahn1} and
Allen-Cahn (AC) Equation for non-conserved systems \cite{Allen1}.
Examples are found in polymer mixtures \cite{Wiltzius1}, alloys
\cite{Shannon1,Gaulin1}, liquid-crystals \cite{Mason1,Chuang1}, and in
cosmology \cite{Laguna1}.

A model that has generated considerable recent interest is the Phase Field
Crystal (PFC) Equation \cite{Elder1,Elder2}, which is a conservative
form of the familiar, non-conserved, Swift-Hohenberg (SH) Equation
\cite{Swift1}. These systems differ from the CH and AC systems in that
the stable phase is periodic.  For SH models, the order parameter is
viewed as capturing the inhomogeneities in a fluid associated with
Rayleigh-B\'{e}nard convection.  In the case of  the PFC model, which is a
simple version of more elaborate density functional theories of
liquid/crystal interfaces \cite{Shih1,Wu1}, the model captures
features at the atomic scale, and thus contains highly detailed
physical information about the system's structure. Such models can
describe many of the basic properties of polycrystalline materials
that are realized during non-equilibrium processing.

The equations of motion governing these non-equilibrium phenomena are
non-linear partial differential equations that cannot generally be
solved analytically for random initial conditions. Therefore, computer
simulations play an essential role in our understanding and
characterization of non-equilibrium phenomena.  The standard Euler
integration is known to be unstable for time step $\Delta t$ above a
threshold fixed by lattice spacing $\dx$ \cite{Rogers1}.  In CH and AC
systems, to maintain an interfacial profile, the lattice spacing must
be smaller than the interfacial width $\xi$, and in PFC and SH
systems, $\dx$ must smaller than the periodicity selected  by
the system. Thus, the Euler update is inefficient, and in practice it
is computationally costly to use to evolve large systems.  Various
computational algorithms \cite{Oono1,Chen1,Zhu1} have been developed
by increasing $\dt$ compared to the simplest Euler discretization.
However, these methods still require a fixed time step, so they
eventually become inefficient.  Recently, unconditionally stable
algorithms \cite{Eyre1,Vollmayr1,Cheng1} were developed to overcome
this difficulty for CH and AC Equations. These algorithms are a class
of stable algorithms free of the fixed time step constraint for
equations with a mix of implicit and explicit terms. While these
algorithms allow for an increasing time step in CH systems as time
proceeds, only a finite effective time step is possible for AC
systems.  A recent study \cite{Cheng2}, based on this unconditionally
stable algorithm, demonstrated analytically  that one can use an accelerated
algorithm $\dt=At^{2/3}$ to drive the CH Equation, with the accuracy in
correlation controlled by $\sqrt{A}$.

In this manuscript we apply this unconditionally stable algorithm to
the PFC and SH Equations (Section $2$). In Section $3$ we establish
the effectiveness of this approach through numerical studies of the
algorithm, demonstrating that the algorithm is both efficient and accurate
for  solving PFC Equation. Finally, in Section $4$ we provide some concluding
remarks.

\section{Unconditionally stable algorithms for PFC Equation}

In this section, we develop a class of unconditionally stable time
stepping algorithms ($\dt$ taken arbitrarily large without the
solution becoming unstable) to the PFC and SH Equations. Although the
main purpose of this section is to study unconditionally stable
algorithms for the  PFC Equation, we include a parallel discussion of the SH
Equation, as the methodology applies to both equations with only
trivial differences.

\subsection{Unconditionally Stable Finite Differences}

Both the PFC and SH Equations start from a free energy functional that
describes the configurational cost of periodic phases in contact with
isotropic phases, and can be expressed as
\begin{equation}
F[\phi]=\int d\vx \left\{\frac{1}{2} \phi\left[r + (1 + \nabla^2)^2\right]\phi 
+ \frac{\phi^4}{4}\right\},
        \label{eq:PFCbasic}
\end{equation}
where the periodic order parameter $\phi({\bf x},t)$ has the wave
number $k_0 = 1$ in equilibrium, and $r<0$  characterizes the  quench depth.
For the PFC equation, $r$ is proportional to the deviation of the
temperature from the melting temperature $T_M-T$.

In the PFC model the order parameter (the density) is conserved, and thus the
equation of motion is in the form of a continuity equation, $\partial
\phi/\p t= -\nabla \cdot {\bf j}$, with current ${\bf j} = - M
\nabla (\delta F/\delta \phi)$, where $M$ is the mobility.  Absorbing
$M$ into the time scale, we obtain the dimensionless form of the PFC Equation
\begin{eqnarray}
\frac{\partial \phi}{\partial t}=\nabla^2 \frac{\delta F}{\delta \phi}
=\nabla^2\left\{\left[r + (1 + \nabla^2)^2\right]\phi +\phi^3\right\}.
        \label{eq:eq_of_motion:PFC}
\end{eqnarray}
For the SH Equation, on the other hand, the order parameter is not
conserved by the dynamics, and its evolution is postulated to have the
the form
\begin{eqnarray}
\frac{\partial \phi}{\partial t}=-\frac{\delta F}{\delta \phi}
=-\left[r + (1 + \nabla^2)^2\right]\phi - \phi^3.
        \label{eq:eq_of_motion:SH}
\end{eqnarray}
Eq. (\ref{eq:eq_of_motion:SH}) has a simple dissipative form, where the
rate of change of $\phi$ is proportional to the gradient (with an an
$L^2$ inner product in functional space) of the free energy.

In order to obtain an unconditionally stable algorithm, we now follow
methods previously developed for the CH and AC Equations
\cite{Vollmayr1,Cheng1}, and work out in some detail how to
semi-implicitly parameterize the equation of motion. We begin by
``splitting'' the linear terms in the equation of motion into
``forward'' and ``backward'' pieces, both for Eq. (\ref{eq:eq_of_motion:PFC}):
\begin{eqnarray}
&&\phi_{t+\dt}+\dt \grad^2
\left[(a_1-1)(r+1)\phi_{t+\dt}+2(a_2-1) \nabla^2 \phi_{t+\dt}
+(a_3-1)\nabla^4 \phi_{t+\dt} \right]
\nonumber \\
=&&\phi_t+\dt \grad^2
\left[a_1(r+1) \phi_t+2a_2 \nabla^2 \phi_t+a_3\nabla^4 \phi_t + \phi_t^3 \right],
        \label{eq:semi_implicit:PFC}
\end{eqnarray}
and for Eq. (\ref{eq:eq_of_motion:SH}):
\begin{eqnarray}
&&\phi_{t+\dt}-\dt
\left[(a_1-1)(r+1)\phi_{t+\dt}+2(a_2-1) \nabla^2 \phi_{t+\dt}
+(a_3-1)\nabla^4 \phi_{t+\dt} \right]
\nonumber \\
=&&\phi_t-\dt
\left[a_1(r+1) \phi_t+2a_2 \nabla^2 \phi_t+a_3\nabla^4 \phi_t + \phi_t^3 \right].
        \label{eq:semi_implicit:SH}
\end{eqnarray}
The constants $a_1$, $a_2$ and $a_3$ control the degree of splitting. In order 
to find the constraints on these parameters that yield an unconditionally stable 
algorithms, a standard von Neumann linear stability analysis on Eq. (\ref{eq:semi_implicit:PFC})
and Eq.  (\ref{eq:semi_implicit:SH}) may be performed. The procedures are quite similar
and the results are identical for these two equations. We will only show the 
details for the PFC model in next subsection.

\subsection{Physical versus numerical instabilities}

As was found in the analysis of Vollmayr-Lee and Rutenberg
\cite{Vollmayr1} for the CH equation, the PFC equation will be
linearly unstable to perturbations for legitimate physical reasons.
Specifically, the isotropic phase $\bar{\phi}$ can be metastable or
unstable to the stable periodic (crystalline) phase \cite{Elder2} if
the system is an undercooled liquid.  This situation (which is
precisely what we are interested in modeling) is established when
$r+3\bar{\phi}^2<0$.  This {\em physical instability} complicates our
standard von Neumann stability analysis, as we wish to predict when
our numerical methods will cause an instability that is unrelated to
the physical instability resulting from the thermodynamic.

We can investigate the physical instability by a linear
stability analysis on the equation of motion Eq. (\ref{eq:eq_of_motion:PFC}). 
We let $\phi=\bar{\phi}+\eta$, where $\bar{\phi}$ is a constant phase and $\eta$ is a small 
perturbation, and linearize the PFC Equation Eq. (\ref{eq:eq_of_motion:PFC}) in $\eta$ to get 
\begin{eqnarray}
\dot{\eta}_t=\grad^2\left[(r+3\bar{\phi}^2)+(1+\nabla^2)^2\right]\eta_t. 
\end{eqnarray}
This can be Fourier transformed to find
\begin{eqnarray}
\dot{\eta}_{\vk,t}=-k^2 \left[(r+3\bar{\phi}^2)+(1-k^2)^2 \right] \eta_{\vk,t}.
\end{eqnarray}
The physical instability for the above equation occurs for
\begin{eqnarray}
\mathbf{r_k}\equiv k^2 \left[(r+3\bar{\phi}^2)+(1-k^2)^2 \right]<0,
        \label{eq:physical_instability}
\end{eqnarray}
which reduces to $r+3\bar{\phi}^2<0$ with $k = 1$ in the stable
phase, as we indicated above. 

Now we can proceed to analyze the {\em numerical stability} and determine the constraints for the 
splitting parameters. We linearize the general step Eq. (\ref{eq:semi_implicit:PFC}) by substituting 
$\phi=\bar{\phi}+\eta$ and get
\begin{eqnarray}
&&\eta_{t+\dt}+\dt \grad^2
\left[(a_1-1)(r+1)\eta_{t+\dt}+2(a_2-1) \nabla^2 \eta_{t+\dt}+(a_3-1)\nabla^4 \eta_{t+\dt} \right]
\nonumber \\
=&&\eta_t+\dt \grad^2\left[a_1(r+1) \eta_t+2a_2 \nabla^2 \eta_t+a_3\nabla^4 \eta_t + 3\bar{\phi}^2 \eta_
t \right],
\end{eqnarray}
The Fourier transform of the above equation results in
\begin{eqnarray}
&&\eta_{\vk,t+\dt}\left[1-\dt k^2 \{ (a_1-1)(r+1)-2(a_2-1) k^2 + (a_3-1)k^4 \} \right]
\nonumber \\
=&&\eta_{\vk,t}\left[1-\dt k^2 \{ a_1(r+1)-2a_2 k^2+a_3 k^4 + 3\bar{\phi}^2 \} \right].
\end{eqnarray}
This can be re-expressed as
\begin{eqnarray}
\eta_{\vk,t+\dt}[1+\dt \mathcal{L}_\vk]=\eta_{\vk,t}[1+\dt \mathcal{R}_\vk].
\end{eqnarray}
Note that $\mathbf{r_k}= \mathcal{L}_\vk-\mathcal{R}_\vk$. While we
want to avoid numerical instability, the physical instability is to be
expected during the dynamics, and will not lead to numerical problems.
But, as we indicated above, both of the instabilities will be captured
by a general von Neumann stability analysis. One manner of dealing
with this is to recognize that a proper unconditionally stable
algorithm will be stable if and only if $\mathbf{r_k}>0$ and should be
unstable if and only if $\mathbf{r_k}<0$.
The von Neumann stability criterion is $|\eta_{\vk,t+\dt}| <
|\eta_{\vk,t}|$.  We can express our restriction on the regime of von
Neumann stability as 
\begin{eqnarray}
\left[1+\dt \mathcal{L}_\vk\right]^2 &>& \left[1+\dt\mathcal{R}_\vk\right]^2
\quad\mathrm{for}\ \mathbf{r_k}>0
\nonumber\\
\left[1+\dt \mathcal{L}_\vk\right]^2 &<& \left[1+\dt\mathcal{R}_\vk\right]^2
\quad\mathrm{for}\ \mathbf{r_k}<0.
\end{eqnarray}
The above inequalities can be rewritten as
\begin{eqnarray}
\mathbf{r_k}\left[2+\dt(\mathcal{L}_\vk+\mathcal{R}_\vk)\right]&>&0
\quad\mathrm{for}\ \mathbf{r_k}>0
\nonumber\\
\mathbf{r_k}\left[2+\dt(\mathcal{L}_\vk+\mathcal{R}_\vk)\right]&<&0
\quad\mathrm{for}\ \mathbf{r_k}<0
\nonumber,
        \label{eq:von_Neumann}
\end{eqnarray}
which, dividing by $\mathbf{r_k}$ can be reduced to a single inequality,
$2+\dt(\mathcal{L}_\vk+\mathcal{R}_\vk)>0$, which implies $0<\mathcal{L}_\vk+\mathcal{R}_\vk$ 
for arbitrarily large $\dt$, and we obtain
\begin{eqnarray}
0<-k^2\left[(r+1)(2a_1-1)+ 3\bar{\phi}^2-2(2a_2-1)k^2+(2a_3-1)k^4\right],
\end{eqnarray}
which can be satisfied using the mode independent restrictions (and $r>-1$)
\begin{eqnarray}
a_1 < \frac{1}{2}-\frac{3\bar{\phi}^2}{2(r+1)}, \ \ \ \ \ \ \ \ \ \ \ 
a_2 \ge \frac{1}{2}, \ \ \ \ \ \ \ \ \ \ \ 
a_3 \le \frac{1}{2}.
        \label{ineq:stability_bounds}
\end{eqnarray}
These are the constraints on the parameters $a_1$, $a_2$ and $a_3$ for
unconditionally stable algorithms for all modes, for quenches in the
range $-1<r<-3\bar\phi^2$. With these choices there is no threshold for 
$\dt$ in order to maintain numerical stability. The
quantity $\dt$ is termed the {\em algorithmic time step}.
We note that unconditional stability does not mean that the user of such algorithms
may simply take as large a time step as is desired. Indeed, to obtain
accurate physical results, there are additional restrictions on how
large $\dt$ may be.

\subsection{Effective time step}

To determine how large a time step we may take, and still maintain an
accurate solution, we calculate the Fourier space ``effective time
step'', as will be described below.
We first note that when $a_1=a_2=a_3=1$, Eq.  (\ref{eq:semi_implicit:PFC})
corresponds to the traditional Euler update
\begin{eqnarray}
\frac{\phi'_{t+\dt}-\phi_t}{\dt_{Eu}}
=\nabla^2\left\{\left[r + (1 + \nabla^2)^2\right]\phi_t +\phi_t^3\right\},
        \label{eq:PFC-Euler-update}
\end{eqnarray}
where $\phi'_{t+\dt}$ denotes the field obtained after an Euler update on a previous 
field $\phi_t$, while we use the unprimed $\phi_{t+\dt}$ to denote the field obtained by 
unconditionally stable algorithm on $\phi_t$ throughout.

We now define the spatial Fourier transform of
$\phi_{\vk,t}=\int d\vx \e^{-i\vk \cdot \vx}\phi_t(\vx)$.
In Fourier space, writing $k^2 \equiv |\vk|^2$, the Euler update becomes
\begin{eqnarray}
\frac{\phi'_{\vk,t+\dt}-\phi_{\vk,t}}{\dt_{Eu}}
=-k^2\left\{\left[r + (1 - k^2)^2\right]\phi_{\vk,t} +(\phi^3)_{\vk,t}\right\},
        \label{eq:PFC-Euler-Fourier}
\end{eqnarray}
where $(\phi^3)_{\vk,t}=\int d\vx \e^{-i\vk \cdot \vx}\phi_t^3(\vx)$.

In Fourier space, the unconditionally stable algorithms Eq. (\ref{eq:semi_implicit:PFC}) 
can be written in a form that is analogous to Eq. (\ref{eq:PFC-Euler-Fourier}):
\begin{eqnarray}
\frac{\phi_{\vk,t+\dt}-\phi_{\vk,t}}{\dt_{eff}^{PFC}(k,\dt)}
=-k^2\left\{\left[r + (1 - k^2)^2\right]\phi_{\vk,t} +(\phi^3)_{\vk,t}\right\},
        \label{eq:PFC-Fourier}
\end{eqnarray}
where we define $k$-dependent effective time step by
\begin{eqnarray}
\dt_{eff}^{PFC}(k,\dt)\equiv\frac{\dt}{1+\dt k^2[(r+1)(1-a_1)+2k^2(a_2-1)+k^4(1-a_3)]}
        \label{eq:dteff:PFC}
\end{eqnarray}
For SH Equation, the effective time step is
\begin{eqnarray}
\dt_{eff}^{SH}(k,\dt)\equiv\frac{\dt}{1+\dt[(r+1)(1-a_1)+2k^2(a_2-1)+k^4(1-a_3)]}.
        \label{eq:dteff:SH}
\end{eqnarray}
$\dt_{eff}(k,\dt)$ is an effective time step for a mode $k$,
corresponding to an algorithmic time step $\dt$. Of particular
interest in the case of periodic systems is the dominant mode (the
lattice spacing in the PFC model), which, for the scaling choices made
in Eq. (\ref{eq:eq_of_motion:PFC}) and Eq. (\ref{eq:eq_of_motion:SH})
is simply $k_0=1$.  Using the parameters employed in the simulations
shown in the next section of $r=-0.025$, $a_1=0.45$, $a_2=0.5$,
$a_3=0.5$, we obtain the dominant effective time step for both
equations
\begin{eqnarray}
\dt_{eff} (k_0,\dt)=\frac{\dt}{1+29 \dt/800}.
        \label{eq:dominant-dteff:PFC}
\end{eqnarray}
As $\dt=\infty$, we obtain the maximum dominant effective time step
$\dt_{eff} (k_0,\infty) = 800/29\approx27.6$.  We see that a large
algorithmic time step $\dt$ does not always translate into a
significant amount of system evolution, as the effective time step
always remains less than 28 for these parameter choices, no matter how
large the algorithmic time step becomes.  Thus, this value provides us
with a useful bound on our exploration of just how large an
algorithmic time step to take, and still obtain accurate results.  For
example, if we take algorithmic time steps that yields an effective
time step $\dt_{eff} (k_0,\dt) = \dt_{eff} (k_0,\infty)/2=400/29$,
then we find $\dt=800/29$. We demonstrate in the next section
that this algorithm, when applied to the PFC Equation, realizes a
significant speedup compared to the traditional Euler algorithm, while
maintaining a controlled level of accuracy.

\section{Numerical results}

\begin{figure}
\begin{center}
\includegraphics[width=13.5cm]{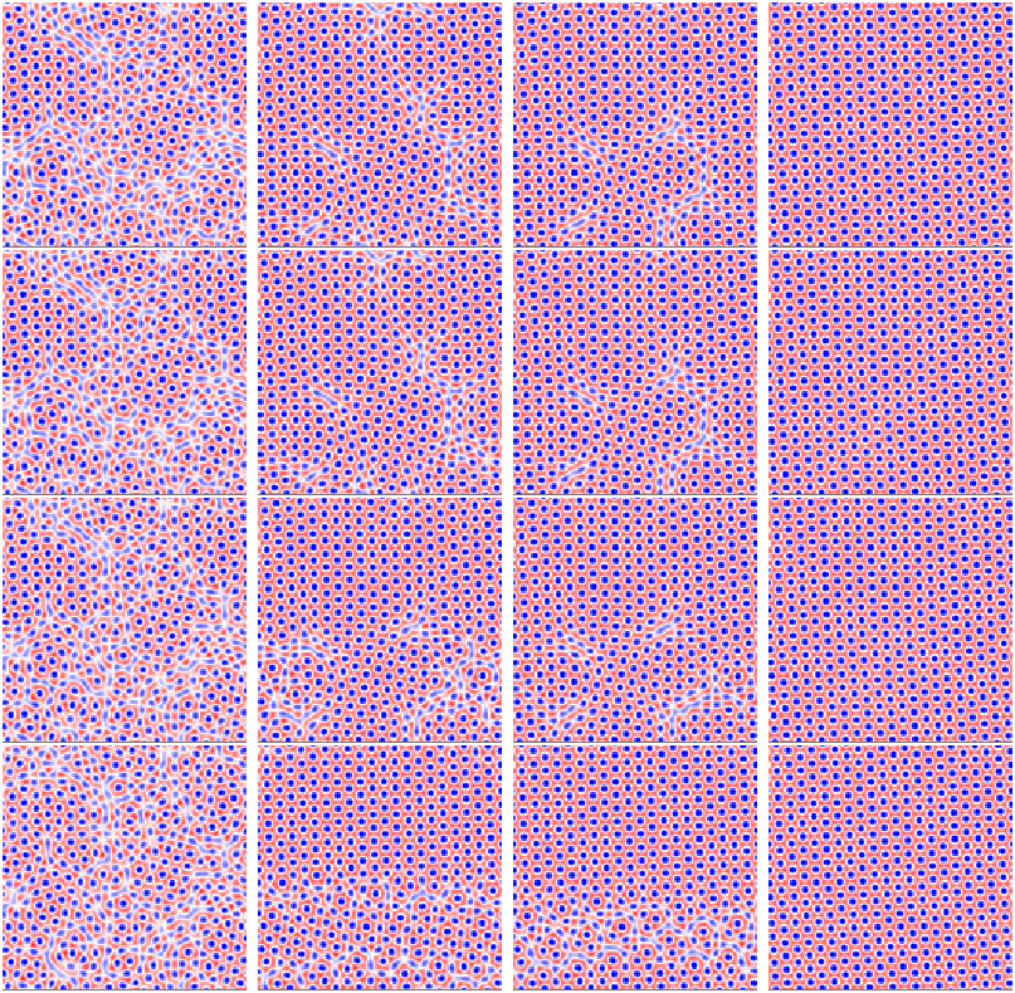}
\caption[Snapshots]{Snapshots of simulations of the PFC model. Time
increases from left to right. The first row shows the field obtained
using the Euler algorithm with $\dt_{Eu}=0.015$. The second to bottom rows show the fields
obtained employing the unconditionally stable algorithms, when using
algorithmic time steps of $\dt=3$, $\dt=10$, and $\dt=30$.}
\label{fig:system_snapshots}
\end{center}
\end{figure}

The simulations were performed in two-dimensions. Fig.
\ref{fig:system_snapshots} shows typical snapshots of simulations for the
PFC model with parameters $\bar{\phi} = 0.07$, $\dx=1.0$, and $L_{sys}
= 128$ with random initial conditions which corresponds to the liquid
state. For comparison, all the simulations start with the same initial
condition. In the Figure white regions indicate $\phi=\bar{\phi}$, red
$\phi=\bar{\phi}+0.2$ and blue $\phi=\bar{\phi}-0.2$. The top row was
obtained using the Euler algorithm $\dt_{Eu}=0.015$ at time steps
$n=30000$, $n=60000$, $n=90000$, and $n=160000$.  The second and lower
rows were obtained using the unconditionally stable algorithm with
(moving down) $\dt=3$, $\dt=10$, and $\dt=30$.  For illustration and
comparison purposes, we show the system snapshots at the same energy
density as the top row --- from left, the energy density $E=0.002374$,
$E=0.002360$, $E=0.002357$, and $E=0.002350$ from the first to fourth
column, respectively.  We immediately see that, for the times and
energies selected, there are no visible differences between the Euler
update simulation and the unconditionally stable algorithm with
$\dt=3$.  However, there are visible differences between the Euler
update and the simulations with $\dt>3$.  We now wish to make these
qualitative observations more quantitative.

\begin{figure}
\begin{center}
\includegraphics[width=12cm]{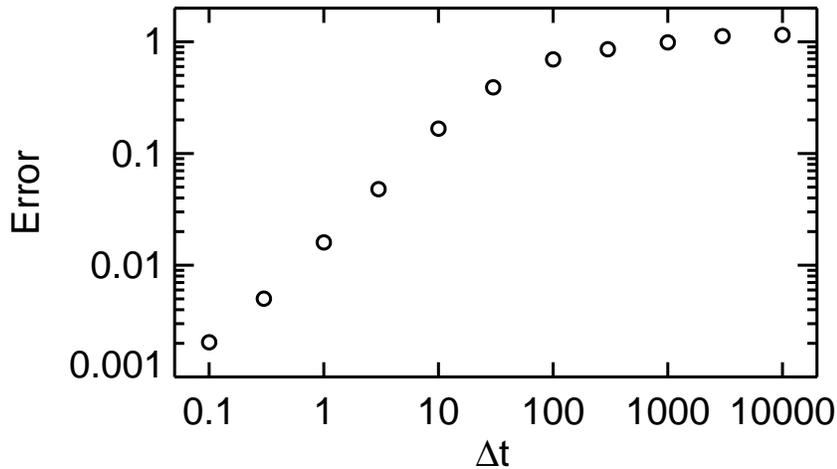}
\caption[Error]{A measure of the error 
  $\sqrt{\langle(\phi_{eu}-\phi_{un})^2\rangle/\langle(\phi_{eu}-\bar{\phi})^2\rangle}$
  versus the algorithmic time step $\dt$.}
        \label{fig:error}
\end{center}
\end{figure}

To study the accuracy, we compare simulations at the same energy
density $E=0.002374$ (the first column in Fig.
\ref{fig:system_snapshots}).  We compute a measure of the error:
$\sqrt{\langle(\phi_{eu}-\phi_{un})^2\rangle/\langle(\phi_{eu}-\bar{\phi})^2\rangle}$,
where $\phi_{eu}(\vx)$ denotes the fields obtained using Euler algorithm
and $\phi_{un}(\vx)$ denotes the fields obtained using the
unconditionally stable algorithm. Fig.  \ref{fig:error} shows a plot
of the error versus a range of algorithmic time steps $\dt$. 
Fig. \ref{fig:error} indicates that, unsurprisingly, the accuracy
increases as we decrease the algorithmic time step $\dt$.  When $\dt
\le 3$, the error is below $5\%$. On the other hand, the error
behavior in Fig. \ref{fig:error} for a large algorithmic time step
tends to saturate, mirroring the saturation in the
effective time step $\dt_{eff}$ for  the dominant mode $k_0=1$.

\begin{figure}
\begin{center}
\includegraphics[width=12cm]{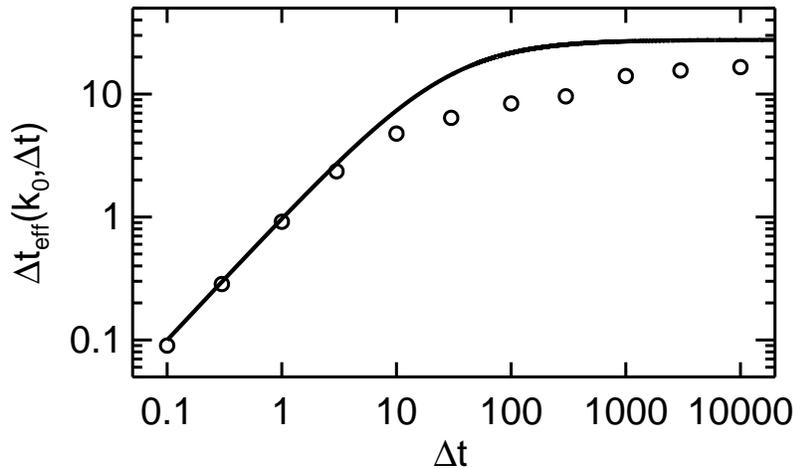}
\caption[Time step ratio]{A comparison between the theoretical dominant effective
time step (solid line) and the numerical estimate of the same quantity (circle).}
        \label{fig:eff_timestep}
\end{center}
\end{figure}

Fig. \ref{fig:eff_timestep} shows a comparison between the dominant
effective time step $\dt_{eff} (k_0,\dt)$ in Eq.
(\ref{eq:dominant-dteff:PFC}) and a numerical estimate of the same
quantity. The numerical estimate is obtained by calculating
$t_{Eu}^{tot}/n_{un}$, where $t_{Eu}^{tot}$ is the total time needed
to reach the final state (a crystalline state without dislocations)
using Euler algorithm and $n_{un}$ is the number of computer steps
needed to reach the same state using unconditionally stable
algorithms. We find good agreement for $\dt \le 3$, while for $\dt>3$,
the separation between the analytic and numerical expressions
increases. While the agreement at small times steps in unsurprising,
the curve provides a useful metric for the optimum algorithmic time
step of $\dt\approx3$, for the chosen parameters.  When $\dt=3$, the
ratio of the number of time steps needed to achieve a particular
energy using the unconditionally stable versus Euler algorithm is
approximately $180$ (the ratio of the dominant mode effective time
step to the Euler time step).  This is a substantial speedup, and
requires minimal analysis to implement the technique.

\section{Conclusions}

In this paper, we have presented an unconditionally stable algorithm 
applicable to finite difference solutions of the the PFC Equation.
We have demonstrated that a fixed algorithmic time step driving scheme
may provide significant speedup, with a controlled level of accuracy,
when compared with Euler algorithm. For the representative parameters
chosen, a speedup of a factor of 180 was obtained.  The analytical
results and the numerical results are consistent with an effective
time step analysis. Although this algorithm allows arbitrarily large
algorithmic time steps, caution is indicated, as  taking too large an
algorithmic time step will yield inaccuracies with little improvement
in the overall speedup of the calculation.  This saturation in the
speedup results from the details of how the system's energy evolution
(and its corresponding microstructural evolution) is governed by the
effective time step, which saturates as the algorithmic time step
increases.  Thus, there is little advantage in too large an
algorithmic time step.  A method for obtaining a reasonable value for
the algorithmic time step $\dt$ is suggested, in which a few test
cases are run with different values of $\dt$ to see which one offers a
good speedup and maintains the desired accuracy.  The analytic form of
the effective time step provides a useful guide for deciding how large
a time step to select when trading off the obtainable speedup versus
the loss of accuracy.

We expect the methodology developed in this paper could find extensive
applications in a wide class of non-equilibrium systems.  For example,
it can be straightforwardly applied to the Swift-Hohenberg Equation, given
its similarity with the Phase Field Crystal model.  Additionally,
the method also will apply to systems where there is a dominant mode
realized at late times, such as is found in diblock co-polymers.
This method should allow researchers to dramatically improve the
computational efficiency associated with modeling the dynamics of
materials systems.

\section{Acknowledgments}

We thank Andrew Reid and Daniel Wheeler for useful discussions.



\end{document}